\documentclass[aprilunum,article,pdftex]{Definitions/apunum} 

\usepackage{natbib}
\usepackage{graphicx,amsmath,amssymb,amstext}

\firstpage{1} 
\pubvolume{1}
\issuenum{1}
\articlenumber{2}
\pubyear{2026}
\copyrightyear{2026}
\datereceived{2026 Apr 1 } 
\dateaccepted{2026 Apr 1} 
\datepublished{2026 Apr 1} 
\dateretracted{2026 Apr 1} 
\pdfoutput=1 
\usepackage{soul}

\usepackage{geometry}
\usepackage{rotating}
\usepackage[normalem]{ulem}

\newcommand{\galaxycoin}{\texttt{GalaxyCoin}} 

\Title{An innovative alternative to traditional funding streams for extragalactic astronomy}

\TitleCitation{GalaxyCoin}

\Author{Stephen M. Wilkins$^{1,\dagger}$, Jack Turner$^{1}$, Connor Sant Fournier$^{2}$, Behnood Bandi$^{1}$, Aswin Vijayan$^{1}$}

\address[1]{$^1$ \hspace{0.2cm} Institute for Extragalactic Finance, Brighton, Commonwealth of Great Britain, Europe, Earth, Sol, Orion–Cygnus Arm, Milky Way, Local Group, Laniakea Supercluster, Pisces–Cetus Supercluster Complex, 
\\ $^{2}$ \hspace{0.2cm} Centre for High Redshift Cryptocurrency, University of Atlam, Msida, Malta, Earth, Sol, Orion–Cygnus Arm, Milky Way, Local Group, Laniakea Supercluster, Pisces–Cetus Supercluster Complex}

\corres{Correspondence: }

\firstnote{If you really feel compelled to email me, you can do so at \href{mailto:s.wilkins@sussex.ac.uk}{s.wilkins@sussex.ac.uk}. However, as \textbf{this is a joke}, correspondence is encouraged only if you wish to thank me for briefly improving your day or to enquire about purchasing \galaxycoin.}  

\abstract{
With traditional sources of funding for astronomical research under increasing pressure, it is timely to explore innovative alternative mechanisms. We therefore introduce \galaxycoin, a novel cryptocurrency whose issuance, validation, and economic evolution are anchored to real astrophysical objects — galaxies. \galaxycoin\ links digital scarcity to observational astronomy by using galaxy catalogues to parametrise token generation, distribution, and long-term supply growth, providing a transparent, immutable, and independently verifiable foundation for the currency. We present the conceptual design of \galaxycoin, highlight its potential advantages over conventional cryptocurrencies, and examine its broader implications for sustainability, trust, and public engagement at the intersection of astronomy, data-driven science, and blockchain technology. A central feature of \galaxycoin\ is that it directly incentivises the discovery and spectroscopic confirmation of galaxies, aligning financial reward with the production of high-quality astronomical data. In terms of monetary design, its supply elasticity lies between that of fiat currencies and fixed-supply cryptocurrencies, making it distinctive in both economic structure and scientific purpose.
}


\begin{document}

\section{Introduction}

Funding for astronomical research and fundamental science more broadly, is increasingly under pressure from shifting government priorities\footnote{\url{https://www.theguardian.com/science/2026/feb/06/uk-scientists-cuts-funding-projects-research-facilities}}\footnote{\url{https://www.nature.com/immersive/d41586-026-00088-9/index.html}}, institutional short-termism, chronic underinvestment, and, in some cases, the whims of despotic autocrats. In order to ensure a sustained flow of resources to support our attempts to understand the Universe, it is essential that the community explores new, unconventional sources of financial support beyond traditional public funding models.

While recent years have seen some success - particularly through the cultivation of philanthropic interest among a small subset of capitalism’s winners\footnote{\url{https://www.skyatnightmagazine.com/news/lazuli-space-observatory-science-paper}} - it is clear that such approaches are neither scalable nor sufficient to sustain the field in the long term. Reliance on sporadic largesse leaves astronomical research vulnerable to economic cycles, personal tastes, and the fickle priorities of individual benefactors\footnote{The authors would like to emphasise that we are \textbf{extremely appreciative} of the financial contributions made by the relatively small number of individuals whose efforts currently sustain key work within our field.}.

To meet these challenges we propose \galaxycoin, a novel cryptocurrency in which each coin is uniquely associated with a real, observable galaxy drawn from verified astronomical surveys. By directly linking digital assets to physical cosmic structures, \galaxycoin\ establishes a transparent, non-arbitrary foundation for value creation while simultaneously introducing a new incentive structure for observational extragalactic astronomy. 

This article is organised as follows. In Section~\ref{sec:core_concepts}, we introduce the core concepts underlying \galaxycoin, including its issuance model and verification framework. Section~\ref{sec:impact} examines the potential impact of \galaxycoin\ on scientific discovery and its role as a novel currency. Additional details on properties (\ref{app:properties}), design choices (\ref{app:design}), and possible extensions (\ref{app:extensions}) are provided in the appendices.

\section{Core concepts}\label{sec:core_concepts}

\begin{figure}[htbp]
    \centering
    \includegraphics[width=0.3\textwidth]{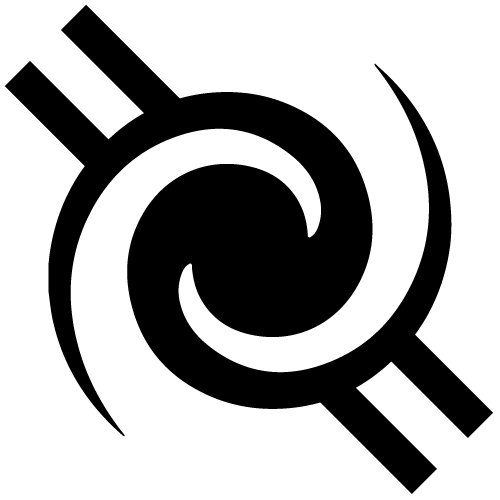} 
    \caption{The proposed symbol for \galaxycoin.}
    \label{fig:symbol}
\end{figure}

\galaxycoin\ (tentative ISO 4217 code \verb|GAL|; proposed symbol shown in Figure \ref{fig:symbol}) is an innovative cryptocurrency designed to support fundamental scientific research while offering an alternative to traditional fiat currencies and conventional cryptocurrencies.

Each \galaxycoin\ is uniquely associated on a one‑to‑one basis with a catalogue entry for a spectroscopically confirmed galaxy. New coins are minted only when a galaxy satisfies a predefined set of observational and validation criteria, typically coinciding with the publication of survey catalogues and the release of the corresponding data - see \ref{app:design} for details of the design principles. In practice, coins will be issued by a dedicated non‑profit entity that administers the minting process and returns the majority of newly minted coins — net of a modest administrative fee — to the originating survey or collaboration for distribution on the open market.

All \galaxycoin\ units carry identical nominal value (see \ref{app:design.value}) and, despite being strictly non‑\emph{fungible} in the sense that each token is unique (see \ref{app:properties.fungibility}), they function as a stable and verifiable medium of exchange within the system. Although individual coins are distinct by design, \galaxycoin\ can nonetheless be used for ordinary commerce: participating vendors and service providers can accept it in exchange for goods and services on the open market.

As discussed in the following section, the design of \galaxycoin\ enables it not only to drive scientific discovery but also to address fundamental limitations of existing fiat and cryptocurrency systems. Nonetheless, \galaxycoin\ has inherent constraints. In particular, its incentive structure primarily targets the study of galaxies and does not directly support other domains of astronomical research\footnote{Though, let's face it, they're not as interesting as observational galaxy formation and evolution are they?}. These limitations can be mitigated by introducing additional, parallel astronomy‑based cryptocurrencies (see \ref{app:extensions}), forming part of a broader ecosystem (\texttt{CosmoCoin}) that collectively supports a wider range of scientific inquiry.

\section{Impact}\label{sec:impact}

\subsection{Implications for scientific discovery}\label{sec:impact.discovery}

A central objective of \galaxycoin\ is to align economic incentives directly with scientific progress in observational astronomy. Because new coins can be minted only through the identification and validation of previously unobserved galaxies, \galaxycoin\ naturally rewards activities that expand the empirical frontier of extragalactic research. In doing so, it incentivises:

\begin{itemize}
    \item The development and deployment of new observational facilities, both ground-based and space-based
    \item Deeper, wider, and higher-resolution surveys capable of probing fainter and more distant galaxy populations
    \item Advances in instrumentation, calibration, data processing pipelines, and survey strategy optimisation
    \item Open data practices, transparent validation criteria, and reproducible scientific workflows
\end{itemize}

Under this model, institutions, collaborations, or consortia responsible for operating telescopes or conducting large surveys may be rewarded through transparent, rule-based mechanisms tied to verifiable discoveries. This creates a novel pathway by which observational success can translate directly into sustained financial support, complementing (rather than replacing) traditional funding streams. Importantly, the incentive structure favours long-term investments in capability and infrastructure, rather than short-term publication metrics.

\begin{figure}[!htbp]
    \centering
    \includegraphics[width=0.95\textwidth]{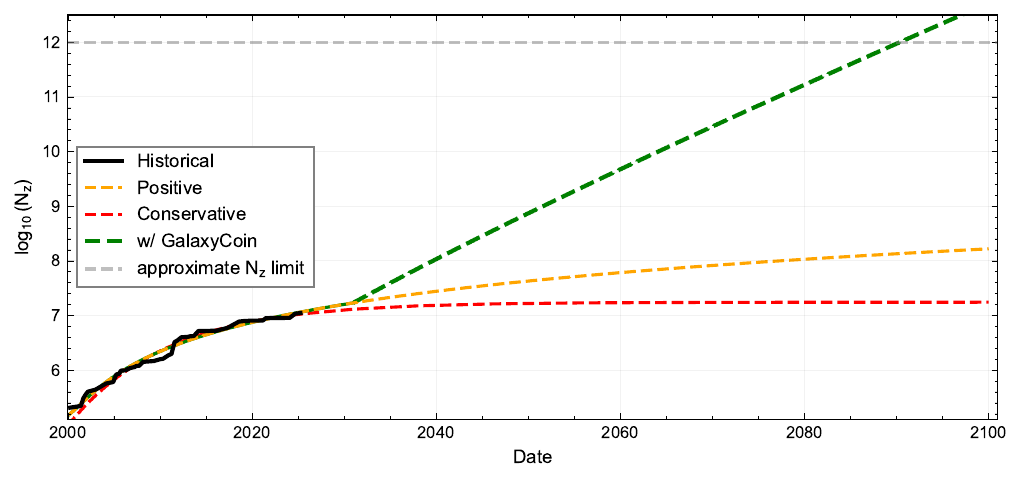} 
    \caption{Projected growth in the number of spectroscopically confirmed galaxies with and without \galaxycoin\ as an incentivising mechanism. The black curve shows the historical trend, based on NASA Extragalactic Database records \citep{Helou_1995}. This is included both for reference and to reassure the reader that at least one part of the figure is based on real data.}
    \label{fig:projection}
\end{figure}

Based on our projections, shown in Figure~\ref{fig:projection}, introducing \galaxycoin\ as an incentive would substantially accelerate the rate of spectroscopic confirmation of new galaxies. Under the \galaxycoin\ scenario we would expect to have spectroscopically confirmed every galaxy in the observable Universe before the end of the 21$^{\rm st}$ century. By contrast, in the absence of \galaxycoin\ the same feat would take at least 6000 years\footnote{Like all good astronomers we simply fit a polynomial to the historical trend and extrapolated it wildly outside its original scope.}, and quite possibly much, much longer — giving future civilisations ample time to wonder what we were up to in the early 2020s.

\subsection{As an actual currency}\label{sec:impact.currency}

Although \galaxycoin’s primary motivation is to incentivise scientific discovery, it also embodies a distinctive monetary structure. Unlike fiat money, whose supply is elastically adjusted by central authorities in response to economic conditions, most cryptocurrencies have a fixed maximum supply released on a predetermined schedule, which can foster deflationary pressure, hoarding and volatility.

\galaxycoin\ occupies a hybrid position: it has a finite upper bound, but its realised supply depends on the number of spectroscopically confirmed galaxies. New coins can only be minted through substantial scientific effort — new facilities, deeper surveys and additional spectroscopic campaigns — rather than by policy decisions or algorithmic schedules. 

This structure gives \galaxycoin\ a degree of transparency unusual for a currency: each unit is backed by an independently verifiable astronomical object recorded in public survey archives. Consequently, supply growth is directly tied to empirical discovery rather than discretionary policy or speculative scarcity. The result is a hybrid monetary model in which scarcity remains meaningful, supply expands through collective investment in knowledge, and trust rests on publicly accessible evidence rather than institutional authority.\footnote{We have been reliably informed by an expert in cryptocurrencies that much of this section is, technically speaking, nonsense.}

\subsection{Public Engagement}\label{sec:impact.pe}

Finally, by tying digital assets to real, observable galaxies, \galaxycoin\ provides an intuitive and engaging entry point into astrophysics and cosmology for non-specialists, transforming abstract discoveries into tangible artefacts. Each token is associated with a specific astronomical object — an individual galaxy with a measurable position, redshift, and observational history — allowing participants to connect directly with underlying scientific data rather than with an abstract financial instrument alone.

This creates a natural bridge between public engagement and research: ownership encourages curiosity about what the object is, how it was observed, how far away it lies, and what it reveals about cosmic history. In principle, participants may become familiar with catalogued galaxies in much the same way that many children — and a surprising number of adults — can effortlessly recite the names, properties, and evolutionary stages of entirely imaginary creatures that are captured, catalogued, and made to fight one another.

In this way, \galaxycoin\ has the potential to turn routine survey outputs into accessible narratives about the expanding Universe, galaxy evolution, and the scale of modern astronomical discovery. A currency backed by galaxies therefore does more than incentivise observation; it also invites broader participation in the scientific imagination.

\section{Conclusions}\label{sec:conclusions}

Astronomical research faces persistent challenges from funding pressures, short-term institutional priorities, and underinvestment. Conventional sources of support — government allocations and philanthropic contributions — are insufficiently scalable or reliable to sustain long-term observational programmes. 

\galaxycoin\ offers a novel mechanism to complement existing funding models by directly linking economic incentives to scientific discovery. Each coin is uniquely associated with a spectroscopically confirmed galaxy, ensuring that new currency enters circulation only through verifiable astronomical research. In doing so, \galaxycoin\ aligns the growth of its monetary base with the expansion of empirical knowledge, directly incentivising scientific research.

Beyond financial support, \galaxycoin\ provides a transparent, verifiable, and engaging interface between the public and the scientific community. By transforming discoveries into tangible, accessible digital assets, it fosters broader participation, education, and trust in astrophysical research.

Overall, \galaxycoin\ demonstrates how emerging technologies can be harnessed to address structural challenges in science funding, embedding curiosity-driven exploration into the design of a self-expanding, verifiable monetary system. As observational capabilities advance, the currency naturally grows alongside humanity’s understanding of the Universe, creating a sustainable, scientifically anchored model for supporting fundamental research.

\funding{This research received no funding whatsoever, obviously. However, if any venture capitalists are interested we've got the Messier Galaxies ready to go.}

\acknowledgments{The authors thank numerous very tolerant colleagues who will be relieved that their lunch time conversations will no longer be dominateed by \galaxycoin.}

\conflictsofinterest{The authors would declare no conflicts of interest but that would obviously be a lie.} 

\section*{Bibliography}

\bibliography{AprilUnum}

\newpage

\appendix
\setcounter{section}{0}

\section{Justification of Design Choices}\label{app:design}

\subsection{Galaxy identification}\label{app:design.identification}

A key challenge in extragalactic astronomy, and for \galaxycoin, is defining what constitutes a galaxy. Observationally, galaxies are typically identified through segmentation, often using every astronomer’s favourite (if dubiously named) software, \texttt{SExtractor} \citep{SExtractor}. Segmentation, however, depends on tunable parameters that can significantly alter the number of detected sources, especially near the sensitivity limit of an observation \citep{Haigh_2021}. Visual inspection is not only mind numbingly boring, but will become completely unfeasible given the increased discovery rate driven by \galaxycoin\ and shown in Figure \ref{fig:projection}.

To address this, we adopt a pragmatic criterion: a \galaxycoin\ corresponds to a galaxy that appears as an isolated object, separated by at least 100 kpc from any previously minted coin\footnote{Assuming some fiducial cosmology.}. This is sufficiently low to allow inclusion of galaxy cluster members \citep{Bird_1995}, but also sidesteps complications from merging galaxies—at least over the expected lifetime of human civilisation \citep{Boylan-Kolchin_2008}. Purely spurious sources must also be addressed. These arise from noise fluctuations dependent on the sensitivity of the observations, so there is no flat criterion that can be applied to mitigate their effect. Instead, a \galaxycoin\ must have an apparent flux greater than the 5$\sigma$ point-source depth of the imaging from which it was identified. This depth will be measured robustly during commissioning phases of new instruments and updated regularly as new observations are processed and instruments degrade over time.

\subsection{The need for spectroscopic confirmation}\label{app:design.spectroscopy}

We restrict eligibility to galaxies with robust spectroscopic confirmation rather than all photometric detections. Spectroscopic follow-up is essential to establish that a source is genuinely extragalactic and not, for example, a nearby brown dwarf enjoying an implausibly successful cosmological disguise \citep{Tu_2025}. More generally, spectroscopy provides reliable redshifts — and therefore distances, luminosities, stellar masses, and chemical abundances — while incentivising an observational activity that remains both indispensable and perpetually oversubscribed.

\subsection{Fixed value}\label{app:design.value}

Each \galaxycoin\ carries the same fixed \emph{face} value, rather than having its worth tied to galaxy properties such as distance or luminosity.

This is partly a matter of simplicity, but also a precaution against accidentally turning the currency into a speculative instrument for cosmologists. If coin values depended on intrinsic luminosity, they would also depend on the adopted cosmological model, meaning that each revision to the expansion history of the Universe could trigger retroactive inflation, deflation, or spirited arguments at conferences. Given current uncertainties \citep{Verde_2019, Roper_2023} and to avoid rewarding whichever cosmologist most recently adjusted $\Lambda$, all coins are assigned an identical and immutable value.

\section{Properties}\label{app:properties}




\subsection{Fungibility}\label{app:properties.fungibility}

Since each \galaxycoin\ is tied to an individual galaxy, it resembles a Non-Fungible Token (NFT) more than a traditional fungible cryptocurrency.

A potential concern is that certain \galaxycoin\ — particularly those representing nearby, scientifically interesting or downright fun \citep{Estrada-Carpenter_2024, van-Dokkum_2024} galaxies—might acquire value beyond their face value, which could limit their usefulness as a currency. This mirrors the situation with some physical coins, where rarity or historical significance gives them value far above their nominal denomination.

However, we expect this to be a minor issue. Galaxies with exceptional additional value likely constitute only a tiny fraction of the total existing supply, and an even smaller fraction of the potential future supply due to observational limitations \citep{Sazonova_2025}.

Consequently, although \galaxycoin\ is strictly non-fungible, this is unlikely to hinder its operation as a practical currency: almost all coins remain identical in face value, with rare exceptions affecting less than 0.01\% of the current supply\footnote{To preserve the stability of \galaxycoin, any scientifically interesting tokens will be discreetly retained by the issuing authority and later marketed at a premium to cover administrative costs.}.

\subsection{Money supply and value}\label{app:properties.supply}

It is useful to briefly consider both the current and potential total supply of \galaxycoin.

At present, the number of spectroscopically confirmed galaxies is approximately $10^7$, according to the NASA Extragalactic Database\footnote{\url{https://ned.ipac.caltech.edu}}, release \verb|35.3.1-TR|. By comparison, the current global money supply, using the broadest (M3) measure, is roughly £$10^{14}$ (£100 trillion)\footnote{\url{https://en.wikipedia.org/wiki/Money_supply}}. Were \galaxycoin\ to instantaneously replace this supply, each token would be worth around £10 million.

While the exact number of galaxies in the Universe is unknown, empirical extrapolations provide meaningful bounds. Integrating observed galaxy stellar mass or luminosity functions over the accessible redshift range—corrected for incompleteness, surface-brightness selection, and cosmic variance—suggests the observable Universe likely contains of order $10^{12}$ galaxies \citep{Conselice2016}. Deep-field observations from the \emph{Hubble Space Telescope} and the \emph{James Webb Space Telescope} indicate that faint, low-mass galaxies dominate the total count, making estimates sensitive to assumptions about galaxy formation efficiency at high redshift. Using this estimate, and assuming that the global economy increases by a factor of $10\times$ in real terms by the time all \galaxycoin\ are minted, this suggests a minimum value of each coin of £1000.

\section{Extensions}\label{app:extensions}

As noted, \galaxycoin\ is not a complete solution to the problem of funding astronomical research and has several limitations. However, some of these limitations can be addressed through extensions of the underlying framework.

\subsection{Multi-wavelength observations}

In its current form, \galaxycoin\ rewards galaxy discovery and spectroscopic confirmation but provides no direct incentive for further observational characterisation. This is a significant limitation, since robust constraints on galaxy properties—such as star-formation rate, dust content, stellar populations, AGN activity, and total mass—require observations across the electromagnetic spectrum, from X-ray to radio \citep{Pacifici_2023}.

Without additional incentives, the system risks reinforcing an existing imbalance in observational astronomy, where initial detections are prioritised over the more resource-intensive follow-up needed for physical interpretation. One possible extension is a family of wavelength-specific parallel currencies, with separate tokens linked to validated observations in distinct spectral regimes (for example X-ray, infrared, radio, or millimetre observations). A single galaxy could then generate multiple complementary tokens, each representing an additional layer of physical information \citep{Conroy_2013}.

Such a framework would better reflect the multi-wavelength nature of modern astrophysics, encourage coordinated observing strategies, and allow the relative value of different observations to evolve with scientific demand. Although this introduces greater complexity in governance and standardisation, it offers a more balanced incentive structure aligned with scientific return.

\subsection{Rare galaxies}

A related challenge concerns rare or extreme systems, such as active galactic nuclei (AGN), hyper-luminous galaxies, or extremely high-redshift \citep{Wilkins2024} galaxies \citep{Lovell2025}, whose scientific value far exceeds their abundance. A uniform \galaxycoin\ model does not naturally capture this asymmetry.

This could be addressed either through dedicated parallel currencies—such as \emph{AGNCoin} — or by weighting rewards according to measurable astrophysical properties such as luminosity, mass, or unusual spectral features. In this way, discoveries of particularly informative systems would generate proportionally greater reward, aligning incentives more closely with scientific importance.

\subsection{Other branches of astronomy}

Although designed for extragalactic astronomy, the framework could be extended to other areas through domain-specific currencies, for example \emph{PlanetCoin} for exoplanets, or \emph{StellarCoin} for stellar astrophysics. Specific categories such as \emph{HabitablePlanetCoin} for potentially habitable worlds and \emph{CowCoin} for bovine based systems \citep{Roper_2022} are also entirely feasible.

Beyond targeted incentives, such currencies could provide a novel mechanism for public engagement, allowing wider communities of users to influence scientific priorities indirectly through participation in a transparent token-based ecosystem. In this sense, \galaxycoin\ and related currencies could function not only as funding tools, but also as a decentralised form of science prioritisation reflecting the diversity of modern astronomy.









\end{document}